%% file: samplepaper.tex
\documentclass[runningheads]{llncs}
\usepackage{graphicx}

\usepackage{booktabs}
\usepackage{algorithm}
\usepackage{algorithmic}
 
 \usepackage{caption}
\usepackage{subcaption}
\usepackage{todonotes}
 \usepackage{mathtools}
 \usepackage[T1]{fontenc}
 
\usepackage{acronym}
\usepackage{amsmath,amsfonts,amssymb}
\usepackage{paralist}
\usepackage{graphicx}

\usepackage{tikz}
\usetikzlibrary{shapes,arrows,fit,backgrounds,automata}
\usetikzlibrary{shapes.geometric,fit}
\tikzstyle{container} = [draw, rectangle, inner sep=0.3cm]
\input{defs.tex}

\begin{document}
\title{On Almost-Sure  Intention Deception  Planning that Exploits Imperfect Observers  \thanks{Research was sponsored by the Army Research Office and was accomplished under Grant
Number W911NF-22-1-0034.}}
\author{Jie Fu\inst{1}\orcidID{0000-0002-4470-2827}}
\authorrunning{J. Fu}
\institute{University of Florida, Gainesville FL 32611, USA   
\email{fujie@ufl.edu}}
\maketitle              % typeset the header of the contribution
\begin{abstract}

Intention deception involves computing a  strategy which   deceives the opponent into a wrong belief about the agent's intention or objective. This paper studies a class of probabilistic planning problems with intention deception and investigates how a defender's limited sensing modality can be exploited by an attacker to achieve its attack objective almost surely (with probability one) while hiding its intention. In particular, we model the attack planning in a stochastic system modeled as a Markov decision process (MDP). The attacker is to reach some target states while avoiding unsafe states in the system and knows that his behavior is monitored by a defender with partial observations. Given partial state observations for the defender, we develop qualitative intention deception planning algorithms that construct attack strategies to play against an action-visible defender and an action-invisible defender, respectively. The synthesized attack strategy not only ensures the attack objective is satisfied almost surely but also deceives the defender into believing that the observed behavior is generated by a normal/legitimate user and thus failing to detect the presence of an attack.    We show the proposed algorithms are correct and complete and illustrate the deceptive planning methods with examples. 
\end{abstract}
\section{Introduction}
%\color{blue}
%All paper {\em submissions} must have a maximum of six pages, plus at most one for references. The seventh page cannot contain {\bf anything} other than references.
%
%
%****Citations within the text should include the author's last name and
%the year of publication, for example~\cite{gottlob:nonmon}.  Append
%lowercase letters to the year in cases of ambiguity.  Treat multiple
%authors as in the following examples:~\cite{abelson-et-al:scheme}
%or~\cite{bgf:Lixto} (for more than two authors) and
%\cite{brachman-schmolze:kl-one} (for two authors).  If the author
%portion of a citation is obvious, omit it, e.g.,
%Nebel~\shortcite{nebel:jair-2000}.  Collapse multiple citations as
%follows:~\cite{gls:hypertrees,levesque:functional-foundations}.
%\nocite{abelson-et-al:scheme}
%\nocite{bgf:Lixto}
%\nocite{brachman-schmolze:kl-one}
%\nocite{gottlob:nonmon}
%\nocite{gls:hypertrees}
%\nocite{levesque:functional-foundations}
%\nocite{levesque:belief}
%\nocite{nebel:jair-2000}****
%
%Deception tactics are featured in strategic decision making in adversarial environments. As more conflicts shifts from military to cyber and physical space, deception and counter-deception have been investigated as an integral part of cyber- and physical defense \cite{janczewski2007cyber} \nocite{}. The understanding of deception will aid in developing effective defense and counter-deception.
%
%\url{https://www.ijcai.org/proceedings/2018/0834.pdf} discussed some results on deception, mainly from AI community.
%\normalcolor

We study a class of intention deception where an attacker   pretends to be a legitimate agent with a benign intention, in order to achieve an adversarial objective without being detected.  Such deceptive behaviors are seen in many   strategic interactions between players with conflicts. For example, in a cyber network, an attacker who carries out advanced persistent attacks can exploit the limitations of the network monitoring capability and operate in a ``low-and-slow'' fashion to avoid detection \cite{chenStudyAdvancedPersistent2014}. Intention deception is the feature of a masquerade attack \cite{salemSurveyInsiderAttack2008}, in which an attacker uses a fake identity to achieve the attack objective without being detected. The success of the intention deception attacks may depend on three factors: First, the defender has incomplete knowledge  about the presence and intention of the attacker; Second, the defender has  imperfect information during an active attack and the lack of information imposes limitations to detect the attack; Third, the attacker can strategically hide his intention behind some  behavior that is deemed normal/legitimate from the defender's viewpoint. %Further, the existence (or the lack) of an intention deception attack strategy can be used to verify the   soundness of a security monitoring system.

 %by exploiting the defender's incomplete and imperfect information.
 The synthesis of deception enables us to assess the vulnerabilities of a defense system against stealthy, intention deception attacks and will aid in developing effective detection methods. Motivated by this need, we develop a formal-methods approach to synthesize intention deception attack strategies.
We model the interaction between the attacker and the targeted system as a \ac{mdp}, in which the attacker has a  stealthy reach-avoid objective, \ie, to reach a set of target states  \emph{with probability one} without being detected or running into any unsafe states. Meanwhile, the defender employs a detection mechanism based on the promise that any attack behavior will sufficiently deviate from a normal user's behavior and thus can be recognized prior to the success of the attack. Our technical approach is motivated by the following question, ``when the defender has an imperfect sensing capability, does the attacker has a stealthy intention deception strategy to achieve the goal, \emph{almost-surely(with probability one)}, while deceiving the defender into thinking his behavior is normal?'' 

Our main contribution is a class of non-revealing intention deception planning algorithms. Non-revealing deception means that the mark (deceivee) is unaware that deception is being used.
Assuming that the state is partially observable to the defender, we  study two cases, referred to as  action visible and action invisible defenders, depending on whether the defender observes the   actions of the agent (either a normal user or an attacker).  The two cases differ  in how the defender refines her belief based on her observations and also what she considers to be rational actions for a normal user.  For each case, we develop  an attack planning algorithm which constructs an augmented \ac{mdp} that incorporates, into the planning state space,   the attacker's information and higher-order information---what the attacker knows that the defender's information. We show that the synthesis of an almost-sure winning, non-revealing intention deception strategy reduces to the computation of an almost-sure winning strategy in the augmented \ac{mdp}. We formally prove the correctness and completeness of the proposed algorithms and illustrate the proposed methods with both illustrative examples and adversarial motion planning examples.

 \paragraph*{Related Work}  %\ac{mdp} modeling of multi-stage attacks  has been widely used in attack-graph based cyber security analysis \cite{lallieReviewAttackGraph2020a,jhaMinimizationReliabilityAnalyses2002,durkotaOptimalNetworkSecurity2015}.
 Intention deception in \ac{mdp}s has been investigated by \cite{karabagDeceptionSupervisoryControl2021a}.  In their formulation,  the optimal deceptive planning is to minimize the Kullback–Leibler (KL) divergence between the (observations of) agent's strategy and (that of) the reference strategy, provided by the supervisor. 
	Comparing to their information-theoretic approaches,  our formulation focuses on qualitative planning with intention deception which exploits the partial observations of the defender/supervisor and does not assume that the attacker/defender knows any specific reference strategy except for an intention of a legitimate user. Such a formulation  is similar to the work in goal/plan obfuscation \cite{mastersDeceptivePathplanning2017}, in which an agent performs path planning in a deterministic environment while making  an observer, equipped with goal recognition algorithms, unable to recognize its goal until the last moment. Recent works on goal obfuscation      \cite{kulkarniUnifiedFrameworkPlanning2019,zhangPlansThatRemain2020,bernardiniOptimizationApproachRobust2020a,kulkarniSignalingFriendsHeadFaking2020} study how a deceptive agent  hides its true goal among finitely many possible decoy goals from an observer. %In \cite{kulkarniUnifiedFrameworkPlanning2019,kulkarniSignalingFriendsHeadFaking2020}, a unified framework is developed so that the agent can make its goal uncertain to its opponent but  legible for its collaborators.
	In \cite{mastersDeceptivePathplanning2017}, the authors study deceptive path planning and define the dissimulation (hiding the real) to occur when the planner ensures the probability of reaching the true goal is smaller than the probability of reaching any other decoy goals. This dissimulation does not extend to almost-sure deceptive planning when the  agent is to ensure that, the probability of satisfying the normal user's objective is one from the defender's partial observations, while also satisfying the attack objective with probability one.
The authors  \cite{bernardiniOptimizationApproachRobust2020a} formulate an optimization-based approach for goal obfuscation given uncertainty in the observer's behaviors. Intention deception is also related to opacity-enforcing control for discrete event system \cite{dubreilSupervisoryControlOpacity2010,linOpacityDiscreteEvent2011,jacobOverviewDiscreteEvent2016,sabooriOpacityEnforcingSupervisoryStrategies2012}, where the supervisor is to design a controller that enforces initial state opacity or  its target language from an intruder with partial observations. % In the supervisory control, researchers have developed opacity-enforcing controller  with respect to state-based predicates or sequential behaviors specified as the target language  \cite{dubreilSupervisoryControlOpacity2010,linOpacityDiscreteEvent2011,jacobOverviewDiscreteEvent2016,sabooriOpacityEnforcingSupervisoryStrategies2012}. 
 However, the discrete-event system models the deterministic transition systems with controllable and uncontrollable actions, \ie, two-player deterministic games, while our work considers planning in \ac{mdp}s, \ie, one-player stochastic games, monitored by a  defender with imperfect observations.
 
 Our problem formulations and solutions are different from existing work in several aspects: 1) we consider probabilistic planning given a partially observable defender, as opposed to   deterministic planning in goal obfuscation and path planning. In deterministic planning problems, it does not matter whether the observer observes the agent's actions. This is not the case with probabilistic planning where the deceptive strategies computed against action-visible and action-invisible defenders require different algorithms; 
 2)  The attacker aims to ensure, with probability one, one of the goals is reached and undetected. In addition, the attacker also needs to satisfy some safety constraints before reaching the goal. This objective is different from goal obfuscation where the agent only plans to reach a single goal, but hides its intention behind multiple decoys; 3) Intention deception is also different from opacity. Opacity in discrete event systems means that from the observer's perspective, the observer cannot infer whether a property is true or false, whileas intention deception is to ensure the observation of a strategy given the attack intention can be mistaken as an observation of a strategy given a benign intention.

The   paper is structured as follows: In the next
section, we introduce the problem formulation. Section ~\ref{sec:main-result} presents the main result for synthesizing almost-surely, non-revealing, intention deception  attack strategies against   two types of defenders (action visible/invisible and state partially invisible). Section~\ref{sec:example} illustrates the attack planning algorithms with examples. Section~\ref{sec:conclude} concludes. %Additional  proofs can be found in the submitted supplementary.

\section{Problem Formulation}
\paragraph*{Notations} Let $\Sigma$ be a finite set of symbols, also known as the \emph{alphabet}. A sequence of symbols $w=\sigma_0 \sigma_1 \ldots \sigma_n$ with $\sigma_i\in \Sigma$ for any $0 \leq i \leq n$, is called a \emph{finite word}. The set $\Sigma^\ast$ is the set of all finite words that can be generated with alphabet $\Sigma$. The length of a word is denoted by $\abs{w}$. %The \emph{empty word}, denoted by $\varepsilon$, is the empty sequence $\Sigma^0$ and $\abs{\varepsilon}=0$. % We denote the set of all $\omega$-regular words as $\Sigma^\omega$ obtained by concatenating the elements in $\Sigma$ infinitely many times. 
%We define the set of nonempty finite words $\Sigma^{+} = \Sigma^{\ast} \setminus \{\varepsilon\}$.  
Given a finite and discrete set $X$, let $\dist X$ be the set of all probability distributions over $X$. Given a distribution $d\in \dist X$, let $\supp(d) = \{x\in X\mid d(x)>0\} $ be the support of this distribution.   

We consider the interaction between an agent (pronoun he/him/his)) and a stochastic environment, monitored by a defender (pronoun she/her) with partial observations. 
There are two classes of agents: One is an attacker  and another is a normal user. 
The stochastic dynamics of the interaction  between the agent and his environment is captured by a Markov decision process (without the reward function but a reach-avoid objective). 

\begin{definition}[Markov decision process (MDP) with a reach-avoid objective]
\label{def:labeled_mdp}
	An  \ac{mdp} with a reach-avoid objective is a tuple $M = \langle S, A, s_0, P , (U, F)  \rangle$, where $S$ and $A$ are finite state and action sets, and $A(s)$ is the set of actions enabled at state $s$, $s_0 $ is the initial state, $P: S \times A \rightarrow \dist{S}$ is the probabilistic transition  function such that $P(s' \mid s, a)$ is the probability of reaching $s' \in S$ given action $a \in A$ is chosen at state $s \in S$. The last component is a pair $(U,F)$ that includes a set $U\subseteq  S$ of \emph{unsafe} states and a set  $F\subseteq S \setminus U$ of \emph{target/goal} states. The \emph{reach-avoid objective} is defined as $\neg U\until F$, where $\until$ is a temporal operator for ``until''. It reads ``Not $U$ until $F$'' and
	represents a set of runs $ \{s_0a_0s_1\ldots \in \plays(M) \mid \exists k \ge 0: s_k\in F\land \forall i<k: s_i \not\in U \cup F \}$ in which  a target state in $F$ is visited at least once  and prior to reach a state in $F$, a state in $U$ is never visited.
\end{definition}The consideration of reach-avoid objectives is motivated by the fact that for  \ac{mdp}s with more complex objectives described by a subclass of temporal logic formula \cite{kupferman2001model}, one can employ a product construction to reduce the planning into an \ac{mdp} with a reach-avoid objective (see Chap.10 of \cite{baier2008principles} and \cite{kwiatkowskaPRISMVerificationProbabilistic2011}). It is also noted that general goal-reaching objectives (also known as reachability objectives) are   special cases of reach-avoid objectives when the unsafe set $U$ is empty. Likewise, the safety objective in which the agent is tasked to avoid a set of states is also a special case of reach-avoid objectives with $F =\emptyset$ \footnote{In temporal logic, the reachability objective is expressed as $\truev \until F$ and the safety objective is expressed as $\neg (\truev\until U)$.}.

A \emph{history} $h= s_0 a_0 s_1 a_1 \ldots s_n$ is an alternating sequence of states and actions, which starts from an initial state and ends in  a state and  satisfies  $P(s_{i+1} \mid s_i,a_i)>0$ for each $ 0 <i < n$. Let $H \subseteq (S\times A)^\ast S$ be the set of  finite histories generated from the \ac{mdp}. A history at time $t$ is a prefix $ s_0a_1s_1a_1\ldots s_{t-1}a_{t-1} s_t $ of a history $h$. 
A \emph{path} $\rho$ is a projection of a history onto the state set $S$, \ie,  for history $h= s_0a_0s_1\ldots s_n$, the path is $\rho= s_0s_1\ldots s_n$.  The last state of a history is denoted $\last(h)$.
The \emph{state-occurrence} function $\Occ: H \rightarrow 2^S$  maps a history to a set of states visited by that history, \ie, $\Occ(s_0a_0s_1a_1\ldots s_n)=\{s_i\mid i=0,\ldots, n\}$.

%A \emph{mixed action} in a state $s\in S$ is a probability distribution over enabled actions $A(s)$. The set of mixed actions in state $s$ is denoted $\dist{A(s)}$. A strategy $\pi$ is a map that to each history $h\in H$, assigns a mixed action $\pi(h)\in \dist{A{s}}$. That is, if $h$ occurs, the strategy selects an action according to the distribution $\pi(h)$. A strategy is \emph{pure}, if it assigns probability 1 on one action to each history. A strategy is \emph{stationary}, if $\pi(h)=\pi(h')$ when $\last(h)=\last(h)$. 

A randomized, finite-memory strategy  is a function $\pi:  H \rightarrow \dist{A}$ that maps a history into a distribution over actions. A randomized, Markov strategy is a function $\pi: S\rightarrow \dist{A}$ that maps the current state to a distribution over actions.  Let $\Pi$ be the set of randomized finite-memory strategies and $\Pi^M$ be the set of randomized, Markov strategies.  We denote $M_\pi = (H, P_\pi, s_0)$ as the stochastic process induced by strategy $\pi \in \Pi$ from the \ac{mdp} $M$, 
where $P_\pi(h a s ' \mid h)  = P(s'|s,a )\cdot \pi(h,a)$ where $s$ is the last state in the history $h$. 
 
 An \emph{event} $E$ is a measurable set of histories. Given an \ac{mdp} and a strategy $\pi$, the probability of events are uniquely defined. We denote by $\Pr(E; M_\pi)$
 the probability of event $E$ occurring in the   stochastic process $M_\pi$. For a reach-avoid objective $\varphi\coloneqq \neg U \until F$, we denote by $\Pr_s(\varphi; M_\pi)$ the probability that $\varphi$ is satisfied, starting from $s$ given the stochastic process $M_\pi$. 
 \begin{definition}[Almost-sure winning strategy and region]
 Given an objective $\varphi$, 	a strategy $\pi:  H \rightarrow \dist{A}$ is almost-sure winning for state $s$ if and only if $\Pr_s(\varphi; M_\pi)=1$.  A set of states from which there exists an almost-sure winning strategy is called the \emph{almost-sure winning region, denoted $\asw$. } Formally, $\asw(\varphi)= \{s\in S\mid \exists \pi \in \Pi: \Pr_s(\varphi; M_\pi) =1\}.$
 \end{definition}
%\paragraph*{Almost-sure winning strategy/region}   A path $\rho=s_0s_1\ldots s_n$ is \emph{winning} for the agent  given the reachability objective $F$ if and only if there exists $i \ge 0$, $s_i\in F$.   A strategy $\pi: S^\ast \rightarrow \dist{A}$ is \emph{almost-sure winning} from the initial state $s_0=s$ if with probability one, a path in the stochastic process induced by this strategy is winning for the agent. 
It is known that for \ac{mdp}s with reach-avoid objectives,  Markov strategies can be sufficient for almost-sure winning.  We provide an algorithm to compute the \ac{asw} Markov strategy in the Appendix.
%	 Formally, let $\Pr(E, M_\pi)$ be the probability of an event $E$ in the   stochastic process induced from $M$ by  strategy $\pi$, we say that $\pi$ is an \ac{asw} strategy if and only if
% 	\[
% 	\sum_{t=0}^\infty \Pr(\Occ(H_t)\cap F\ne \emptyset, M_\pi)=1.
% 	\]  
% 	where $H_t=S_0A_0S_1\ldots S_t$ is  a history of  $t$ time steps. 
%	The set of states starting from which there exists an \ac{asw} strategy given the reachability objective $F$ is called the agent's \ac{asw} region, denoted $\asw(F)$. It is noted that an \ac{asw} strategy may not be unique.

For simplicity, we introduce a function called the ``post'', defined as follows: For a subset $X\subseteq S$ of states and an action $a\in A$,   $\post(X,a)=\{s\in S\mid \exists s' \in X, P(s|s',a)>0\} $, which is the set of states that can be reached by action $a$ with a positive probability from a state in the set $X$.
 
\noindent \textbf{The attacker's and user's objectives}  
The attacker's reach-avoid objective is given by $
\varphi_1 \coloneqq \neg U_1\until F_1$ and  the user's objective is given  by $\varphi_0 \coloneqq \neg U_0 \until F_0$, where $U_i, F_i\subseteq S$, $i=1,2$. %To make the problem nontrivial, we consider the case when $F_0\cap F_1 =\emptyset$, because otherwise, the attacker can hide his intention perfectly by carrying out a strategy to reach $F_0\cap F_1$. 
%
%
%\begin{definition}[\ac{ltl} \cite{manna2012temporal}] 
%    Given a set of atomic propositions $\calAP$, an \ac{ltl} formula is defined inductively as follows:
%    \[
%        \varphi \coloneqq p \mid \neg \varphi \mid \varphi \land \varphi \mid  \Next \varphi \mid \varphi \until \varphi,
%    \]
%    where $p\in \calAP$ is an atomic proposition. The temporal operator $\Next$ is called the ``next'' operator. The formula $\Next \varphi$ means that the formula $\varphi$ will be true in the next state. The temporal operator $\until$ is called the ``until'' operator. The formula $\varphi_1 \until \varphi_2$ means that $\varphi_2$ will become true in some future time steps, and before that $\varphi_1$ holds true for every time step. Using the until operator, we define an operator $\Eventually$ (read as eventually) as follows: $\Eventually \varphi = \top \until \varphi$ where $\top$ is unconditionally true. The formula $\Eventually \varphi$ is true if $\varphi$ holds in some future time. 
%\end{definition}

We refer an attacker and a normal user as an \emph{agent} and consider the setting when the agent's activities are monitored by a supervisor/defender, who has however imperfect observations. Specifically, the defender's observation is defined by his observations of states and actions of the agent:
\begin{definition}[State-observation function] The state-observation function of the defender is $\obs_S: S  \rightarrow 2^S$ that maps a state $s$ to a set $\obs_S(s)$ of states that are observation equivalent to $s$. 
For any $s'\in \obs_S(s)$, it holds that $\obs_S(s')=\obs_S(s)$.
\end{definition}

The observation equivalence relation defined by $\obs_S$ forms a partition over the state set $S$. For the observation function to be properly defined, it holds that $s\in \obs(s)$ for any $s\in S$. 
The following standard assumption is made.
\begin{assumption}
For any state $s'\in \obs_S(s)$ that $s\ne s'$, the sets of enabled actions satisfy $A(s)=A(s')$. 
\end{assumption}
 That is, if two states are observation equivalent, then the  agent have the same set of actions enabled  from these two states. Similarly,  We define the action-observation function.

\begin{definition}[Action-observation function]
The action-observation function of the defender $\obs_A: A\rightarrow 2^A$   maps an action $a$ to a set of actions observation equivalent to $a$. 
For any $a'\in \obs_A(a)$, it holds that $\obs_A(a')=\obs_A(a)$.
\end{definition}
In the scope of this work, we 
  restrict to two special cases with action observation:  1) The defender is \emph{action invisible} if she cannot observe which action is taken by the agent. In this case, 
$\obs_A(a)=  A$---that is, all actions generate the same observation and are indistinguishable.
2) Otherwise, the defender is said to be \emph{action visible} and $\obs_A(a)=\{a\}$. These two classes of action observation functions are commonly considered in qualitative analysis of partially observable stochastic systems \cite{chatterjeePartialobservationStochasticGames2011}.

We combine the state observation and action observation functions into a single observation function $\obs:A\times S\rightarrow 2^A \times 2^S$ such that $\obs(a,s) = (\obs_A(a),\obs_S(s))$. The  observation function extends to histories of the \ac{mdp} recursively, that is, given a history $h=s_0a_0s_1a_1\ldots s_n$, the defender's observation history is $\obs(h)= \obs_S(s_0)\obs_A(a_0)\ldots \obs_S(s_n)$.
\section{Main Results}
\label{sec:main-result}

We are interested in the following question: Given an attacker who aims to achieve an attack objective in the \ac{mdp}, does he has a strategy to do so while the defender, who observes his behaviors, mistakes the attacker as a normal user? If the answer is affirmative, the strategy used by the attacker is called \emph{non-revealing, intention deception strategy}. We investigate the problem for both action-invisible and action-visible defenders, with partial state observations.

\subsection{Nonrevealing intention deception attack planning against action-visible defender}
\label{subsec:action-visible-single}

The following information structure is considered for the attacker, the normal user, and the defender.

\paragraph*{Information structure: An action-visible defender against  an attacker with perfect information}
\begin{inparaenum}[1)]
    \item Both the attacker and the defender know the user's  objective $\varphi_0$. 
    \item The defender does not know the attacker's  objective $\varphi_1$. 
    \item The defender has partial observations over states, defined by the state-observation function $\obs_S: S\rightarrow 2^S$ and is \emph{action  visible}.
\end{inparaenum}

\begin{definition}[Observation-equivalent histories and strategies]
\label{def:obs-equivalent}
  Two histories $h, h'$ are observation equivalent if and only if $\obs(h)=\obs(h')$. 
  Two strategies $\pi_0,\pi_1$ are \emph{qualitatively observation-equivalent}  if and only if the following condition holds: 
  \begin{multline*}
  \forall h \in H,   \Pr(h; M_{\pi_0}) >0  
  \implies \\ \exists h' \in H, (\obs(h') = \obs(h)) \text{ and } 
  \Pr(h'; M_{\pi_1}) >0,
  \end{multline*}
  and vice versa.
\end{definition}
Intuitively, for an attacker's strategy to be qualitatively observation equivalent to a user's strategy, it means that when the attacker carries out his attack strategy, for any history $h$ with a nonzero probability to be generated, there is an observation-equivalent history $h'$  that has a nonzero probability to be generated by the user's strategy. 

The reason for us to define the observation equivalence in this manner is because for qualitative planning with  the almost-sure winning objective, the  planner only needs to reason about whether a history  has a positive probability to be sampled but not the exact probability.  
 This is due to the following property of almost-sure winning strategy.

\begin{proposition}
	Let $\pi$ be an \ac{asw} strategy in the \ac{mdp} $M = \langle S,A, s_0, P, (U, F) \rangle$ with reach-avoid objective, for any history $h$ such that $\Pr(h, M_\pi)>0$ and $\Occ(h)\cap F= \Occ(h)\cap U = \emptyset$, 	$\Pr (\exists h' \in (A\times S)^\ast: \Occ(h h')\cap F\ne \emptyset \text{ and } \Occ(hh')\cap U = \emptyset;  M_\pi )=1$.
\end{proposition}
In words, any finite history with a positive probability to be sampled in $M_\pi$ will have a suffix that visits a state in $F$ while avoiding $U$. It directly follows from the definition of almost-sure winning strategy (see also \cite{baierPrinciplesModelChecking2008}).

 Based on Def.~\ref{def:obs-equivalent}, we now define formally the attacker's non-revealing, intention-deception, \ac{asw}  strategy.
\begin{definition}
\label{def:asw-nonreveal}
An attack strategy $\pi_1^\ast$ is called \emph{non-revealing intention deception, \ac{asw}  strategy} if and only if it satisfies the following conditions:
\begin{inparaenum}[1)]
    \item $\pi_1^\ast$ is qualitatively observation-equivalent to an \ac{asw} strategy $\pi_0$ of the user in the \ac{mdp} $M$ with the   objective $\varphi_0$.
\item $\pi_1^\ast $ is almost-sure winning in the \ac{mdp} $M$ with the objective $\varphi_1$. 
\end{inparaenum}
\end{definition} 
Note that the user may have multiple (potentially infinite, randomized) \ac{asw} strategies. Thus, instead of following a specific user's \ac{asw} strategy, the attacker would like to know what set of actions will be rational for the user in any \ac{asw} strategy. This brings us to define the set of permissible actions.
%\aknote{Are ASW strategies unique? If not, the proof may not be straightforward. Does $\pi_1^*$ depend on which $\pi_0^{\asw}$ is selected by user?}
%\jf{No. the user has multiple ASW strategy, as long as the attacker' strategy is observation equivalent to one of them. the attacker cannot be detected.}

\begin{definition}[Permissible Actions]	
	Given an \ac{mdp} $M$ with a reach-avoid objective $\varphi \coloneqq \neg U \until F$, for $s\in \asw(\varphi)$, an action $a\in A(s)$ is \emph{permissible} if there exists an \ac{asw} strategy $\pi: S\rightarrow \dist{A}$   such that $\pi (s,a)>0$. 
\end{definition}

\begin{lemma}Given an \ac{mdp} $M$ with a reach-avoid objective $\varphi = \neg U\until F $, for $s\in \asw(\varphi)$, action $a \in A(s)$ is permissible only if   $\post(s,a) \subseteq \asw(\varphi)$.
\end{lemma}
In words, a permissible action ensures the agent stay within his almost-sure winning region. %Note that the agent cannot ensure to reach the target set of $\varphi$ by selecting  permissible actions. Instead, he must ensure making progress towards $F$ with a positive probability in finitely many steps and only permissible actions are allowed. (The supplementary material provides  the computation of \ac{asw} region and permissible actions. )

Let 
$\allowed: S\rightarrow 2^A $ be a function defined by $\allowed(s) = \{a\in A(s)\mid \post(s,a)\subseteq \asw(F)\}$.
\begin{proposition}
\label{prop:property-of-ASW}
	Given   \ac{mdp} $M$ with a reach-avoid objective $\varphi \coloneqq \neg U \until F$,  a Markov strategy $\pi: \asw(\varphi) \rightarrow \dist{A}$ that satisfies $\supp(\pi(s))=\allowed(s), \forall s\in \asw(\varphi)$, is almost-sure winning. 
\end{proposition}
The proof is in the Appendix. With this proposition, we will allow the attacker to ``hide'' his attack strategy behind a user's \ac{asw} strategy by calculating the permissible actions for the user.
We denote by $\allowed_0$ the function that maps a state to a set of user's permissible actions.

Next, we construct an \ac{mdp} with augmented state space for synthesizing a non-revealing, intention deception, \ac{asw} strategy for the attacker.

%\todo[inline]{The permissible strategy may need to consider the case different given safety and reachability objectives. For reachability objective, an action is permissible if it has a positive probability to ``make progress'' and has a probability one to stay safe. }

\begin{definition}[Attacker's deceptive planning against action-visible defender]
\label{def:action-visible-plan}
Given the \ac{mdp} $M$ and two objectives $\varphi_0 = \neg U_0 \until F_0, \varphi_1 = \neg U_1\until F_1$ capturing the normal user and attacker's  objectives, respectively. The following augmented \ac{mdp} is constructed:
\[
\tilde M = \langle  S\times 2^S, A, \tilde P, (s_0, \obs_S(s_0)),  (\tilde U_1, \tilde F_1) \rangle ,
\]
where 
\begin{itemize}
\item $S\times 2^S$ is a set of states. A state  $(s,B)$, where  $B\subseteq S$, consists of a state $s$ in the original \ac{mdp} $M$ and a belief $B\subseteq S$ of the defender about the state from past observations;
\item The transition function $\tilde P$ is constructed as follows.   For $(s,B)\in S\times 2^S$ and $B\ne \emptyset$,
if there exists $s^o\in B$, $a\in \allowed_0(s^o)  $,  then 	   for each state $s'' \in \post(s,a)$,  let $\tilde P((s'',B')|(s,B),a) = P(s''|s,a)  $ where \begin{equation}
	\label{eq:belief_update} B' =  \bigcup_{s^o\in B: a\in \allowed_0(s^o)} \{s' \mid   s' \in \post(s^o,a) \text{ and } \obs_S(s') = \obs_S( s'')\}.\end{equation} 
For any $s\in S$, any $a\in A(s)$, let
$\tilde P((s,\emptyset)|(s,\emptyset),a)=1$.
 \item $(s_0, \obs_S(s_0))$ is the initial augmented state. 
\item $\tilde U_1 =\{(s, X)\mid s\in U_1, X\subseteq S, X\ne \emptyset\} \cup \{(s, \emptyset)\mid s\in S\}$ and $  \tilde F_1 = F_1\times (2^S \setminus \emptyset)$ are  the set of unsafe states and  the set of target states in the augmented \ac{mdp} for the attacker, respectively. The attacker's objective is a reach-avoid objective $\neg \tilde U_1\until \tilde F_1$.
\end{itemize}
\end{definition}
The transition function in $\tilde M$ is well-defined because given the next state, the observation is determined and thus the new belief given the state and action observations is determined. 

The second component of an augmented state represents the defender's belief given past observations. The probabilistic transition function is understood as follows:
Consider the defender holds a belief  $B$ about the current state, and observes that action $a$ is taken. 
Then, for each state $s^o\in B$, the defender  considers if $a$ is a permissible action at $s^o$. If the answer is yes, then the defender will compute the set of reachable states from $s^o$ given $a$ using $\post(s^o,a)$ and removes a subset of reachable states that are not consistent with the observation generated by the actually reached state $s''$.
In this process, 
the defender   eliminates some states in $B$ for which the action $a$ is not permissible, under the rationality assumption for the normal user. 

Finally, the objective is defined using an unsafe set $\tilde U_1$. By definition, if any state in $\tilde U_1$ is reached, then either \begin{inparaenum}[1)]
    \item the attacker reaches an unsafe state $s\in U_1$ in the original \ac{mdp} and thus violate the reach-avoid objective; or \item the defender's belief given past observation becomes an empty set. This means the defender knows   the observed behavior is not a normal user's behavior. Thus,  $(s, \emptyset)$ is a sink state. Once a sink   $(s, \emptyset)$ is reached,  the attack is revealed.
\end{inparaenum}

% \begin{remark}
% A key observation is that though the defender believes that the agent will stay within the user's winning region $\asw(F_0)$, the attacker can reach a state $s\notin \asw(F_0)$ but have the defender believes that the current state is $s' \in \asw(F_0)$  which is observation-equivalent to $s$. We shall see it in the example later. %That is the reason we don't restrict the attacker to stay within the user's winning region $\asw_0$. 
% \end{remark}

%\aknote{If $\obs_S$ partitions $S$, the condition under which attacker can stay stealthy behind user's strategy depends largely on $\obs_S$. Can we determine this condition on $\obs_S$? E.g. (conjecture.) If $\obs_S(s) \cap F_1 =\emptyset$ holds for every state $s \in F_0$, then attacker must reveal attack at least on the last step. Can a stronger condition/strategy be derived that enforces attacker to reveal early?}   
%\jf{this is a much harder design problem. maybe considered in the next paper.}

Given the augmented \ac{mdp} $\tilde M$ and the   attack objective $\neg \tilde U_1\until \tilde F_1$, we can   compute an \ac{asw} strategy $\tilde \pi_1:S\times 2^S\rightarrow \dist{A}$ such that for any state $(s,B)$ where $\tilde \pi_1$ is defined, by following $\tilde \pi_1$, the attacker can ensure a state in $\tilde F$ is reached with probability one. It is observed that $\tilde \pi_1$ cannot be reduced to a Markov strategy in the original \ac{mdp} because for the same state, say $s$, but different belief states $B\ne B'$, $\tilde \pi_1(s,B)$ may not equal $\tilde \pi_1(s,B')$. As the size of belief states is finite, the attack strategy $\tilde \pi_1$ can be viewed as a finite-memory strategy in the original \ac{mdp}, with memory states as the defender's belief states. The finite-memory strategy induces a stochastic process $M_{\tilde \pi_1}$ from the original \ac{mdp}. 

\begin{theorem}
\label{thm:action-visible}
 An intention deception, non-revealing, almost-sure winning strategy for the attacker is the \ac{asw} strategy in the augmented \ac{mdp} $\tilde M$ with reach-avoid objective $\neg \tilde U_1\until \tilde F_1$. 
\end{theorem}
\begin{proof}
We show that the \ac{asw} strategy $\tilde \pi_1: S\times 2^S\rightarrow \dist{A}$ obtained from the augmented \ac{mdp} is qualitatively observation-equivalent to an \ac{asw} strategy for the user, as per Def.~\ref{def:asw-nonreveal} and Def.~\ref{def:obs-equivalent}.

Consider a history $h  = s_0a_0s_1a_1\ldots s_n $ which is sampled from the stochastic process $M_{\tilde \pi_1}$ and satisfies $s_i \notin F_1 \cup U_1$ for $0\le i <n$ and  $s_n \in F_1$ . The history is associated with a history in the augmented \ac{mdp}, \[\tilde h =  (s_0, B_0)a_0(s_1,B_1)a_1\ldots (s_n,B_n),\] where $B_0=\obs_S(s_0)$ is the initial belief for the defender. Due to the construction of the augmented \ac{mdp}, for all $0\le i \le n-1$, $B_i\ne \emptyset$.  Otherwise if there exists $i =1,\ldots, n-1$, $B_i=\emptyset$, then $\tilde F_1$ will not be reached because  state $(s,\emptyset) $ for any $s \in S$ is a sink state.
By the definition of qualitatively observation-equivalence, we only need to show that there exists a history $h' =s_0'a_0s_1'a_1 \ldots s_n'$ where $s_i'\in B_i,i=0,\ldots, n$, such that $\Pr(h'; M_{\pi_0})>0$ where $M_{\pi_0}$ is the Markov chain induced by a user's \ac{asw} strategy $\pi_0$ from the \ac{mdp} $M$.

Suppose, for any state-action sequence 
$h'  = s_0'a_0s_1'a_1\ldots s_n' $ where   $s_i'\in B_i$ for $0\le i \le n$, it holds that $\Pr(h'; M_{\pi_0})=0$. When $\Pr(h'; M_{\pi_0})=0$, there are two possible cases: First case:  there exists some $i\ge 0$ such that for all $s\in B_i$, $a_i\notin \allowed_0(s)$; second case: for any $s\in B_i$ that $a_i\in \allowed_0(s)$, there does not exist $s'\in B_{i+1}$ such that $P(s'|s,a_i)>0$.
%\aknote{See if $\obs_S$ partitioning $S$ has any effect on this argument.}
 
Clearly, the first case is not possible because in that case, no transition will be defined for action $a_i$ given the belief $B_i$
(See \eqref{eq:belief_update}) and thus $B_{i+1} =\emptyset$. It contradicts the fact that $B_{i+1}\ne \emptyset$.  In the second case, it holds that $\post(s, a_i)\cap B_{i+1}=\emptyset$ for any $s \in B_i$. However, by construction, we have $B_{i+1}  = \bigcup_{s\in B_i, a_i \in \allowed_0(s)} \post(s,a_i) \cap \obs_S(s_{i+1})$ and thus  \\$B_{i+1} \subseteq \bigcup_{s\in B_i, a_i \in \allowed_0(s)} \post(s,a_i)$. If $\post(s, a_i)\cap B_{i+1} =\emptyset$ for any $s\in B_i$, then it is only possible that $B_{i+1} =\emptyset$, which is again a contradiction. 

Combining the analysis of these two cases, we show that there exists such a history $h'$ such that $\obs(h)=\obs(h')$ and $\Pr(h'; M_{\pi_0})>0$. As the history $h$ is chosen arbitrarily, it holds that for any history sampled from $M_{\tilde \pi_1}$, we can find an observation-equivalent history that has a non-zero probability to be generated in the Markov chain $M_{\pi_0}$.
\end{proof}

\subsection{Non-revealing intention deception against action-invisible defender}

We anticipate that the attacker   have more advantages in non-revealing intention deception if the defender is action-invisible. To this end, we construct the attacker's planning problem against an action-invisible defender. The information structure is similar to that for the action-visible defender, except the defender cannot observe the attacker/user's actions.
\begin{definition}[Attacker's deceptive planning against action-invisible defender]
\label{def:action-invisible-plan}
Given the \ac{mdp} $M$ and $\varphi_0, \varphi_1$ capturing the user's and attacker's  objectives, respectively. Assuming that the defender is action invisible and partially observes the states, the attacker constructs the following augmented \ac{mdp}:
\[
\widehat M = \langle  S\times 2^S, A, \widehat P, (s_0, \obs_S(s_0)),   (\widehat U_1, \widehat F_1) \rangle ,
\]
where 
\begin{itemize}
\item $S\times 2^S$ is a set of states. A state  $(s,B)$, where  $B\subseteq S$, consists of a state $s$ in the original \ac{mdp} $M$ and a belief $B\subseteq S$ of the defender about the state from past observations;
\item The transition function $\widehat P$ is constructed as follows.   For $(s,B)\in S\times 2^S$, If $\bigcup_{s^o\in B} \allowed_0(s^o)\ne \emptyset$, then for each action $a\in \left(\bigcup_{s^o\in B} \allowed_0(s^o) \right) \cap A(s)$,  let $\widehat P((s'',B')|(s,B),a) =  P(s''|s,a)   $ where the new belief $B'$ is computed as 
\begin{multline}
	\label{eq:action-invisible-belief-update}
B' =  \bigcup_{s^o\in B} \{s' \mid \exists  a^o\in \allowed_0(s^o). \; s' \in \post(s^o,a^o) \\ \text{ and } \obs_S(s') = \obs_S( s'')\}.
\end{multline}
Note that   action $a^o$ does not necessarily equal action $a$. 
%	(\bigcup \{s^o \mid \exists s \in B, \exists a' \in \allowed_0(s), s^o \in \post(s,a')\} )\cap \obs_S(s')$. 

  If $\bigcup_{s^o\in B} \allowed_0(s^o)=\emptyset$,  $\widehat P((s,\emptyset)|(s,B),a) =  1   $, and $\widehat P((s,\emptyset)|(s,\emptyset),a)=1$ for any $a\in A(s)$. 
 \item $(s_0, \obs_S(s_0))$ is the initial augmented state.

\item $\widehat U_1 =\{(s, X)\mid s\in U_1, X\subseteq S, X\ne \emptyset\} \cup \{(s, \emptyset)\mid s\in S\}$ and $  \widehat F_1 = F_1\times (2^S \setminus \emptyset)$ is a set of unsafe states and target states in the augmented \ac{mdp} for the attacker. The attacker's objective is a reach-avoid objective $\neg \widehat U_1\until \widehat F_1$.
\end{itemize}
\end{definition}

The main difference between the augmented \ac{mdp} $\tilde M$ and $\widehat M$ is in the definition of the transition function: For action-invisible defender, its belief update   takes into account of all possible and permissible actions  for a normal user, and the observation of that reached state. For action-visible defender, the belief update will be employ the action observation.  
% The transition function is understand as follows. Given that the current state is $s$ and the defender's belief is $B$, the attacker takes an action $a$ which cannot be observed by the defender. The next state $s'$ is reached probabilistically according to the original \ac{mdp} dynamics. The defender receives an observation $\obs_S(s')$ of the state $s'$ and reasons about the belief as follows: Suppose that the state $s^o\in B$ was the actual state before the attacker takes an action, the defender thinks that an action $a^o\in \allowed_0(s^o)$ shall be taken by a rational user. According to this action effect, the current state can be any state in the set $\post(s^o,a^o)$ that is \emph{consistent} with his observation $\obs_S(s')$. Therefore, the defender will consider all possible states in $B$ and all possible allowed actions to calculate the possible next state $B'$, taking into account of his observation. 

\begin{theorem}
\label{thm:action-invisible}
 An intention deception, non-revealing, almost-sure winning strategy for the attacker is the \ac{asw} strategy in the augmented \ac{mdp} $\widehat M$ with the reach-avoid objective $\neg \widehat U_1 \until \widehat F_1$. 
\end{theorem}

The proof is similar to that of the action visible defender case and thus omitted. The reader can find the complete proof in the Appendix.

It is not difficult to see the following statement holds.
\begin{lemma}
  Any non-revealing intention-deception \ac{asw} attack strategy against an action-visible defender is  a non-revealing intention-deception \ac{asw} attack strategy against an action-invisible defender.
\end{lemma}
\begin{proof}[Sketch of the proof]
If the defender cannot detect a deviation from the user's strategy given the observations of states and the  actions, he cannot detect such a deviation when he has no action information.
\end{proof}

\begin{remark}
It is noted that the qualitative objective for almost-sure winning does not consider the cost/time it takes for the attacker to achieve the objective. The \ac{asw} strategy  is not unique (See Proposition~\ref{prop:property-of-ASW}). Therefore, additional planning objectives can be considered to compute an \ac{asw} strategy, such as,  minimizing the expected number of steps or total cost to reach the target set while avoiding unsafe states. 
\end{remark}

\subsection{Complexity analysis}
The time complexity of \ac{asw} strategy in an \ac{mdp} is polynomial in the size of the \ac{mdp} \cite{gimbertComputingOptimalStrategies2011,baierPrinciplesModelChecking2008}. Therefore, the computation of nonrevealing intention deception \ac{asw} strategy for either action-visible or action-invisible defender is polynomial in the size of the augmented \ac{mdp} and thereby exponential in the original \ac{mdp} due to the subset belief construction. This is expected for qualitative analysis of partially observable stochastic games on graphs \cite{chatterjee2010qualitative,chatterjeePartialobservationStochasticGames2011}.

\section{Examples}
\label{sec:example}

This section includes a number of illustrative examples and examples with real-world security applications to demonstrate the   methods.
\subsection{Illustrative examples}

\input{illustrative_ex.tex}

\subsection{Intention deception     planning against a security monitoring system}
\input{robot_ex.tex}

\section{Conclusion and future work}
\label{sec:conclude}
We develop a formal-method approach to synthesize an almost-sure winning strategy with intention deception in stochastic systems. Such a strategy exploits the defender's partial observations over state-action space. With the proposed methods, it is possible to assess whether a defender's   monitoring capability is vulnerable to such intention deception attacks.
This work has several future extensions: First, the insight from synthesizing intention deception can lead to possible counter-deception. For example, the defender can design her observation function/sensing modality to eliminate the existence of such deceptive attack strategies; Second, the paper investigates an attacker to hide his identity as one normal user. If there are multiple normal user behaviors, we hypothesize that the attacker can have more advantages by exploiting the defender's uncertainty about the type of normal user whose behavior is being observed. Third, the relation between qualitative intention deception and quantitative intention deception using information-theoretic approaches \cite{karabagDeceptionSupervisoryControl2021a} may be established. Instead of finding the almost-sure nonrevealing intention deception strategy, it may be practical to also compute an  intention deception strategy that has a low probability of revealing the true intention. 
Lastly, the synthesis of intention deception can be extended to general observation functions for practical applications in cyber-physical systems. 

\appendix
\input{appendix.tex}

\bibliographystyle{splncs04}

\bibliography{refs}
\end{document}

%% file: defs.tex
\newif\ifuseboldmathops
\newif\ifuseittextabbrevs
\useboldmathopstrue   % comment out to use mathbb
\ifuseittextabbrevs

	\newcommand{\ie}{{\it i.e.}}

\else

	\newcommand{\ie}{i.e.~}

\fi

\ifuseboldmathops
\else

\fi

\ifuseboldmathops

\else

\fi

\ifuseboldmathops

\else

\fi

\ifuseboldmathops

\else

\fi

\newcommand{\truev}{\mathsf{true}}

\newcommand{\falsev}{\mathsf{false}}

\newcommand{\until}{\mbox{$\, {\sf U}\,$}}

\newcommand{\abs}[1]{\lvert#1\rvert}

\newcommand{\supp}{\mbox{Supp}}

\newcommand{\Occ}{\mathsf{Occ}}

\newcommand{\last}{\mathsf{Last}}

\newcommand{\prog}{\mathsf{Prog}}

\newcommand{\plays}{\mathsf{Plays}}
\newcommand{\obs}{\mathsf{DObs}}

 \acrodef{mdp}[MDP]{Markov Decision Process}
\acrodef{pomdp}[POMDP]{Partially Observable MDP}
 \acrodef{tl}[TL]{Temporal Logic}
  \acrodef{vm}[VM]{Virtual Machine}
  \acrodef{mtd}[MTD]{Moving Target Defense}
    \acrodef{pctl}[PCTL]{Probabilistic Computation Tree Logic}   
    \acrodef{asw}[ASW]{Almost-Sure Winning}

        \acrodef{sdn}[SDN]{software-defined networking}

\newtheorem{assumption}{Assumption}

\newcommand{\dist}[1]{\mathcal{D}(#1)}

\newcommand{\post}{\mathsf{Post}}

\newcommand{\asw}{\mathsf{ASW}}

\newcommand{\allowed}{\mathsf{Allowed}}

%% file: illustrative_ex.tex
\begin{figure}[ht!]
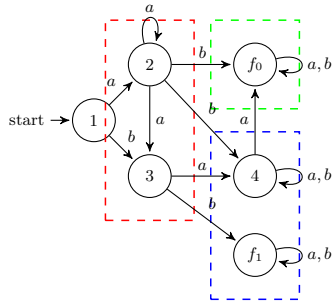

	\centering 
	\include{mdp-ex}
	\caption{An \ac{mdp} without reward. Each box includes a set of observation-equivalent states.}
	\label{fig:mdp-ex}
	\end{figure}

We first use an  illustrative example, shown in Fig.~\ref{fig:mdp-ex} to show the construction of deceptive policies. In this example, the attacker is to reach   $F_1 = \{f_1\}$ and the user is to reach set $F_0= \{f_0\}$, where the unsafe sets $U_0$ and $U_1$ are empty.  The edges are labeled with actions that triggers a probabilistic transition. For example, if action $a$ is taken at state $2$, then   the next state is either $2$ or $3$, with some positive probabilities. For clarity, we omitted the exact probabilities on these transitions in Fig.~\ref{fig:mdp-ex}.
States $f_0, f_1$ are sink states.  
First, we synthesize the user's \ac{asw} region (see Algorithm in the supplement) to compute the permissible actions for each state: $\allowed_0(1)=\{a,b\}$, $\allowed_0(2)=\{a,b\}$, $\allowed_0(3)=\{a\}$, and $\allowed_0(4)=\{a\}$.  It can be shown that at each state, if the user takes an allowed action with a positive probability, he can reach $f_0$ with probability one.
Clearly, if the attacker does not play intention detection, he will select action $b$ from state $3$ to reach $f_1$ directly, but also reveal himself to the defender, had the defender knows that the current state is $3$ and is action-visible.

\paragraph*{Deceiving an action-visible defender} We construct the augmented \ac{mdp} according to Def.~\ref{def:action-visible-plan} and solve the non-revealing deceptive  attack policy.
\begin{figure}[ht!]
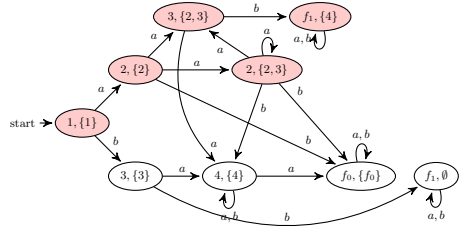

	\centering 
	\include{augmdp-ex}
	\caption{The augmented \ac{mdp} for attack planning against an action-visible defender.  The red notes are the attacker's \ac{asw} region for nonrevealing intention deception. }
	\label{fig:augmdp-ex}
\end{figure}
The transitions in the augmented \ac{mdp} in Fig.~\ref{fig:augmdp-ex} is understood as follows: Consider for example the state $(2,\{2\})$ which means the current state is $2$ and the defender's belief is $\{2\}$. When the defender observes action $a$ being taken, he can deduce that the next state can be either $2$ or $3$. The transitions $(2,\{2\})\xrightarrow{a} (2,\{2,3\})$ and $(2,\{2\})\xrightarrow{a} (3,\{2,3\})$ describe the probabilistic outcomes to reach states $2$ and $3$, and the defender's belief is updated to $\{2,3\}$. At state $(3,\{2,3\})$, if the action $b$ is taken, then the only possible next state is $f_1$. However, because the defender does not know the exact state (3 in this case), she observes $\{4,f_1\}$ and thinks the behavior is still normal if the current state is 2. This is because  state $4$  is still in the \ac{asw} region for the user. 

However,  if action $b$ is observed at the state $(3,\{3\})$, then with probability one, the next state is $(f_1,\emptyset)$, which is a sink state. The reason for the second component to be $\emptyset$ is because action $b$ is not permissible from $3$ for the user. Similarly, when action $b$ is taken from $(3,\{2,3\})$, the next state is $(f_1,\{4\})$ because $b\in \allowed_0(2)$, $\post(2,b) \cap \obs_S(f_1) = \{f_0,4\}\cap \{4,f_1\} =\{4\}$. This is an interesting case where the defender's belief no longer contains the true state. The reason for this to happen is that the defender assumes that only permissible actions will be taken by a user.

Next, we compute the \ac{asw} policy for the attacker in the augmented \ac{mdp} given the goal to reach $(f_1, \{4\})$. The policy is the following: $\pi((3,\{2,3\}))= b$, $\pi(2,\{2,3\})= \pi(2,\{2\})=a$, and $\pi((1,\{1\}))=a$. The red nodes are the attacker's \ac{asw} region for intention deception. This policy ensures the state $f_1$ will be reached with probability one and from the defender's partial observation, the behavior is possible for a normal user. For example, a history $h$ can be $1\xrightarrow{a} 2 \xrightarrow{a}2 \xrightarrow{b} 4 $, which is consistent with the defender's observation and the user's policy. It is a prefix of a history $1\xrightarrow{a} 2 \xrightarrow{a}2 \xrightarrow{b} 4 \xrightarrow{a} f_0$ that reaches $f_0$. The true history $h^\ast$ is $1\xrightarrow{a} 2 \xrightarrow{a}3 \xrightarrow{b} f_1$, which is observation equivalent to $h$.

\begin{figure}[ht!]
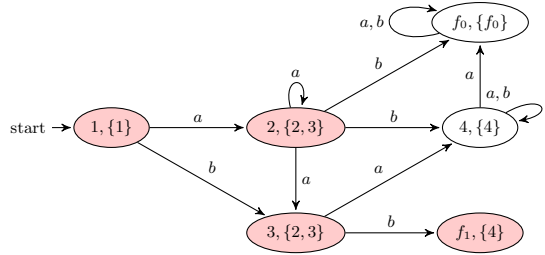

	\centering 
	\vspace{-3ex}
	\include{augmdp-noact-ex}                \vspace{-3ex}
	\caption{The augmented \ac{mdp} for attack planning against an action-invisible defender. The red nodes are the attacker's \ac{asw} region for intention deception.}
	\vspace{-3ex}
	\label{fig:augmdp-noact-ex}
\end{figure}
\paragraph*{Deceiving an action-invisible defender}
We construct the augmented \ac{mdp} according to Def.~\ref{def:action-invisible-plan}. The transition in the augmented \ac{mdp}  in Fig.~\ref{fig:augmdp-noact-ex} is understood as follows: Consider for example the state $(1,\{1\})$,  $\widehat P((3,\{2,3\} ) \mid (1,\{1\}), b) >0$, where 
$2$ is included in the belief is because the action $a$ is allowed at 1 for user and with that action, $2$ can be reached with a positive probability and $\obs_S(2) = \obs_S(3)$.
% At the state $(3, \{2,3\})$, if action $a$ is taken, then $\widehat P((4,\{4\} ) \mid (3,\{2,3\}), a) >0$ where $B' =\{4\}$.   
Note that   states $(2,\{2\})$ and $(3,\{3\})$ are not reachable in the action-invisible case because the defender cannot observe if action $a$ or $b$ is taken at the initial state.  One of the attacker's \ac{asw} deceptive policy is $\hat \pi_1((1,\{1\}))=b$ and $\hat \pi_1((3,\{2,3\}))=b$. When the attacker commits to this policy, the observation obtained by the defender is $\{1\}, \{2,3\}, \{4\}$, for which the following history $1\xrightarrow{b} 3\xrightarrow{a} 4$ is possible and may be generated by an \ac{asw} user policy as from state $4$, the user can still reach $f_0$ with probability one by taking action $a$. It is interesting to see that this non-revealing \ac{asw} policy for the attacker would be revealing if the defender is action-visible.

%% file: mdp-ex.tex
\begin{tikzpicture}[->,>=stealth',shorten >=1pt,auto,node distance=2 cm, scale =0.7,transform shape]
	
	% Configuration 1
	\node[state, initial] (1) {$1$};
	\node[state] (2) [above right of=1, node distance=1.5cm] {$2$};
	\node[state] (3) [below right of=1,node distance=1.5cm] {$3$};
	\node[state] (f0) [right of=2] {$f_0$};
	\node[state] (4) [right of=3] {$4$};
	\node[state] (f1) [below   of=4,node distance=1.5cm] {$f_1$}; 
	\node [container,fit=(2) (3),draw=red,dashed,line width=0.2mm] (container) {};
	\node [container,fit=  (f0),draw=green, dashed,line width=0.2mm] (container) {};
	\node [container,fit=(4) (f1), draw=blue, dashed,line width=0.2mm] (container) {};
	
	\path (1) edge node {$a$} (2)
	(1) edge node {$b$} (3)
	(2) edge[loop above] node {$a$} (2)
	(2) edge node {$a$} (3)
	(2) edge node {$b$} (4)
	(3) edge node {$b$} (f1)
	(2) edge node {$b$} (f0)
	(3) edge node {$a$} (4)
	(4) edge node {$a$} (f0)
	(4) edge [loop right] node {$a,b$} (4)
	(f1) edge [loop right] node {$a,b$} (f1)
	(f0) edge [loop right] node {$a,b$} (f0)
	
	;
	
\end{tikzpicture}

%% file: augmdp-ex.tex
\begin{tikzpicture}[->,>=stealth',shorten >=1pt,auto,node distance=2.5cm, scale =0.5,transform shape]
	
	% Configuration 1
	\node[ellipse, draw, initial,fill=red!20] (1) {$1,\{1\}$};
	\node[ellipse, draw] (2) [above right of=1, node distance=2cm,fill=red!20] {$2,\{2\}$};
	\node[ellipse, draw] (3) [below right of=1,node distance=2cm] {$3, \{3\}$};

	\node[ellipse, draw] (2-23) [right of=2, node distance=3.5cm,fill=red!20] {$2, \{2,3\}$};
	\node[ellipse, draw] (3-23) [above right of=2,node distance=2cm,fill=red!20] {$3, \{2,3\}$};
	\node[ellipse, draw,fill=red!20] (f1-4f1) [right of=3-23,node distance =3.5cm]
	{$f_1,\{4\}$}; 
	\node[ellipse, draw] (4) [right of=3]
	{$4,\{4\}$};
	% \node[ellipse, draw] (4-4f0) [right of=2-23, node distance=3.5cm]
	% {$4,\{f_0\}$};
	\node[ellipse, draw] (f0-4f0) [right of=4,node distance=3.5cm]
	{$f_0,\{f_0\}$};
	\node[ellipse, draw] (f1) [right    of=f0-4f0, node distance=2cm]
	{$f_1,\emptyset$};
	%  \node[ellipse, draw] (4-empty) [right of=f0-4f0,node distance=3.5cm] {$(4, \emptyset)$};
	\path (1) edge node {$a$} (2)
	(1) edge node {$b$} (3)
	(2) edge node {$a$} (2-23)
	(2) edge node {$a$} (3-23)
	(2) edge node [pos=0.8]{$b$} (f0-4f0)
	% (2) edge [bend left=90] node {$b$} (4-4f0)
	(3) edge [bend right] node {$b$} (f1)
	(3) edge node {$a$} (4)
	(4) edge [loop below] node   { $a,b$} (4)
	(4) edge node {$a$} (f0-4f0)
	%  (4-4f0) edge [bend left=60] node {$a$} (f0-4f0)
	%  (4-4f0) edge [loop below] node {$a,b $}  (4-f0)
	(3-23) edge node {$b$} (f1-4f1)
	(2-23) edge  node[pos=0.2] {$b$} (f0-4f0)
	(2-23) edge node[pos=0.2] {$b$} (4)
	(2-23) edge node {$a$} (3-23)
	(3-23) edge[bend right] node[pos=0.9] {$a$} (4)
	(2-23) edge[loop above] node {$a$} (2-23)
	(f1-4f1) edge[loop below] node [pos=0.8,left,yshift=2] {$a,b$} (f1-4f1)
	%  (4) edge [loop below] node {$b$} (4)
	(f0-4f0) edge[loop above] node {$a,b$} (f0-4f0)
	(f1) edge[loop below] node {$a,b$} (f1)
	;
	
\end{tikzpicture}

%% file: augmdp-noact-ex.tex
\begin{tikzpicture}[->,>=stealth',shorten >=1pt,auto,node distance=2cm, scale =0.7,transform shape]
	
	% Configuration 1
	\node[ellipse, draw, initial,fill=red!20] (1) {$1,\{1\}$};
	\node[ellipse, draw,fill=red!20] (2-23) [right of=1, node distance=3.5cm] {$2, \{2,3\}$};
	\node[ellipse, draw,fill=red!20] (3-23) [below of=2-23] {$3, \{2,3\}$};
	% \node[ellipse, draw] (4) [right of=3-23,node distance =3.5cm] {$4,\{4\}$};
	\node[ellipse, draw,fill=red!20] (f1-4f1) [right of=3-23,node distance=3.5cm]
	{$f_1,\{4\}$};
	\node[ellipse, draw] (4-4f0) [right of=2-23, node distance =3.5cm]
	{$4,\{4\}$};
	\node[ellipse, draw] (f0-4f0) [above  of=4-4f0] 
	{$f_0,\{f_0\}$};  
	\path (1) edge node {$a$} (2-23)
	(1) edge node {$b$} (3-23) 
	% (4) edge node   { $a$} (4-4f0)
	%  (4) edge [loop below] node   { $a,b$} (4)
	(4-4f0) edge [in=10,out=30,loop] node  [pos=0.1] { $a,b$} (4-4f0)
	% (4) edge [bend right=90] node[pos=0.9] {$a$} (f0-4f0)
	(4-4f0) edge node {$a$} (f0-4f0)
	(f0-4f0) edge [loop left]node {$a,b$} (f0-4f0)
	(3-23) edge node {$b$} (f1-4f1)
	(2-23) edge node {$b$} (f0-4f0)
	(2-23) edge node {$b$} (4-4f0)
	(2-23) edge node {$a$} (3-23)
	(3-23) edge node {$a$} (4-4f0)
	(2-23) edge[loop above] node {$a$} (2-23)
	;
	
\end{tikzpicture}

%% file: robot_ex.tex
\begin{figure}[ht!]
	\centering
	\begin{subfigure}[b]{0.4\textwidth}
		\includegraphics[width=0.5\textwidth]{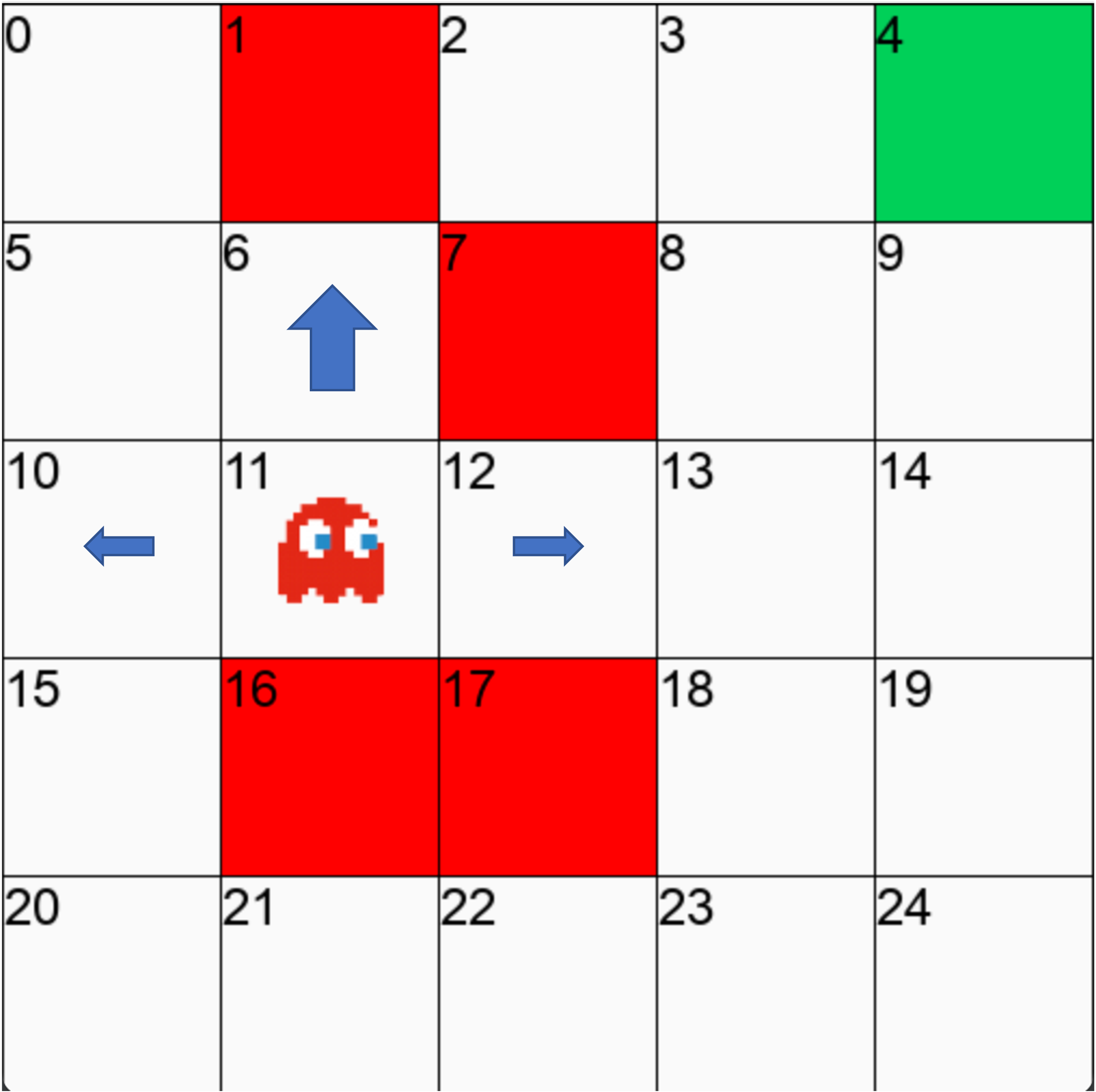}
		\caption{
		\label{fig:gridworld}}
	\end{subfigure}
	\begin{subfigure}[b]{0.4\textwidth}
		\input{sensor_scheduler}    \caption{
		\label{fig:sensor}   }
	\end{subfigure}
\caption{(a) A stochastic gridworld.  (b) The probabilistic sensor scheduler. }
\end{figure}

\begin{figure}[ht!]
	\begin{subfigure}[b]{0.3\textwidth}
		\begin{tabular}{ll}
			State  & Coverage    \\
			1 &  $\{0,1,2,3,4\}$  \\
			2 &    $\{3,8,13,18,23\}$ \\
			3& $\{15,16,17,18,19\}$  \\
			4 & $\{1,6,11,16,21\}$
		\end{tabular}
\caption{	\label{tbl:config-a} }
	\end{subfigure}
	\begin{subfigure}[b]{0.3\textwidth} 
		\begin{tabular}{ll}
			State & Coverage    \\
			1 &  $\{5,6,7,8,9\}$  \\
			2 &    $\{3,8,13,18,23\}$\\
			3& $\{5,6,7,8,9\}$  \\
			4 & $\{3,8,13,18,23\}$
		\end{tabular} \caption{  \label{tbl:config-b}    }
	\end{subfigure}
\begin{subfigure}[b]{0.3\textwidth} 
	\begin{tabular}{ll}
		State & Coverage    \\
		1 &  $\{0,1,2,5,6,7\}$  \\
		2 &    $\{22,23,24\}$\\
		3& $\{0,1,2,5,6,7\}$  \\
		4 & $\{22,23,24\}$
	\end{tabular}  \caption{ \label{tbl:config-c} }
\end{subfigure}
\caption{Sensor configurations (a), (b), (c).}
\label{fig:sensor_configs}
\end{figure}
We now use another example to demonstrate the applications of the methods in security applications. Consider a stochastic gridworld  environment shown in Fig.~\ref{fig:gridworld} in which an agent (a user or an attacker, shown as the red ghost) is to achieve some reach-avoid objective. 
The agent has four actions: ``up'', ``down'', ``left'', ``right''. For each cell $s \in S$, after taking an action, the robot transits to the \emph{correct} cell with probability $p$ and slips to unintended cells with probability $1-p$. As illustrated in the figure: When the agent decides to go ``up'', with probability $p$, it reaches the intended cell 6, and with   $\frac{1-p}{2}$ probability to reach 10 and 12, respectively.  Red cells are obstacles. If the agent runs into an obstacle, it gets stuck. The environment is surrounded by bouncing walls. If the agent takes an action and hits a wall, then the agent stays in the original cell.   

The environment is monitored by a sensor network with a randomized schedule shown in Fig.~\ref{fig:sensor}. 
 The monitoring system runs a finite-memory probabilistic scheduler with four states: $Q= \{1,2,3,4\}$ with the probabilistic transition dynamics shown in Fig.~\ref{fig:sensor}.  At every time step, the scheduler has a $50\%$ chance of staying in the same state for next time step and $50\%$ chance of switching to a different state. 
 At each state $q\in Q$, the sensor coverage is given by $\gamma(q)$. 
 We consider two types of   sensors: 1) type (B) is a Boolean sensor that returns $\truev$ if the agent's state is in the coverage. For example, if $\gamma(1) = \{5,6,7,8,9\}$, then under the sensor state $1$,   if the agent is in one of the cells in $\gamma(1)$, the sensor returns $\truev$. Otherwise, the sensor returns $\falsev$.
2) Type (P) is a precise range sensor that returns the exact agent's state if the agent's state is in the coverage. For example, if $\gamma(1) = \{5,6,7\}$ and the agent is in state $5$ and the sensor is in state $1$, then the sensor returns $\{5\}$. 
Different sensor types and  coverages as well as different finite memory schedulers will generate numerous sensor configurations for the defender to obtain the state information of the agent. We consider three sensor configurations (different coverages but the same scheduler) in Table.~\ref{tbl:config-a}, \ref{tbl:config-b}, and \ref{tbl:config-c}. The configurations $(b), (c)$ have essentially two sensors that are  switched on and off probabilistically. We assume that the defender knows the agent's state   at the beginning of the game. 

\noindent \textsl{User's and attacker's intentions: }The user's objective is to avoid obstacles and state $4$ until the goal $0$ is reached, \ie, $\varphi_0\coloneqq \neg U_0 \until F_0$ where $U_0 =\{1,7,16,17,4\}$ and $F_0=\{0\}$. The attacker's objective is to avoid obstacle until the goal $\{4\}$ is reached, \ie, $\varphi_1\coloneqq \neg U_1\until F_1$ where $U_1= \{1,7,16,17\}$ and $F_1=\{4\}$. Given that the two objectives conflict with each other, $\varphi_0\land \varphi_1= \falsev$, if the defender has complete observations over the state space, regardless action visible or not, the defender should be able to detect the attacker. 

To illustrate the exploitability of such a sensor network, we sample a number of sensor configurations and  investigate when the attacker has an intention deception strategy in each of the configurations. The result is concluded in Table~\ref{tbl:results} \footnote{The computation is performed in a MacBook Pro with 16 GB memory and Apple M1 Pro chip. The computation time is the total time taken to compute the attacker's  intention deception  \ac{asw} region in the augmented \ac{mdp}.} . 
The sensor configuration id is shown as $\mbox{(x)-Y-Z}$ where $x\in \{a,b,c\}$ represents to the sensor coverage configurations in Fig.~\ref{fig:sensor_configs}, $Y\in \{B,P\}$ represents if the sensor type is Boolean (B) or precise (P), and $Z\in \{I, V\}$ represents if action is invisible  (I) or visible (V).
 We show the sets of initial states from which the attacker has an \ac{asw} deception strategy, regardless of the sensor states. These states are referred to as the winning initial states. Comparing $\mbox{(a)-B-I}$ with  $\mbox{(b)-B-I}$, it shows that $\mbox{(b)-B-I}$ includes more initial states that are deceptively winning for the attacker, as well as a larger ratio of states in the augmented \ac{mdp} belonging to the almost-sure winning region. In this example, when the defender is action visible, the winning initial state sets for both configuration $(a),(b)$ are empty.  
 
 In the case of sensor configuration $(c)$, when the defender is action invisible,  the attacker has a non-revealing, intention deception, \ac{asw}  strategy for a set of initial states  for both Boolean and precise sensor types. The set of winning initial states is smaller when the sensors are precise. However, when we have both action visible and precise sensor type, the winning initial state set is empty for $\mbox{(c)-P-V}$. The non-empty \ac{asw} region in the case of $\mbox{(c)-P-V}$ includes these  states in the augmented \ac{mdp} at which the defender's belief   is not a singleton.
 
\begin{figure}[ht!]
	\centering
	\begin{subfigure}[b]{0.45\textwidth}
		\includegraphics[width= \textwidth]{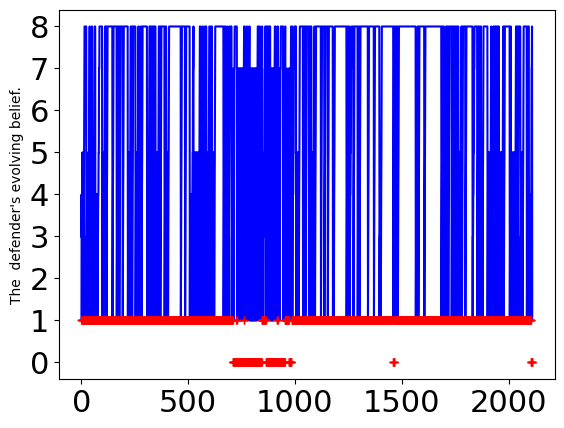}
		\caption{Sensor (a),B,I}
	\end{subfigure}
	\begin{subfigure}[b]{0.45\textwidth}
		\includegraphics[width= \textwidth]{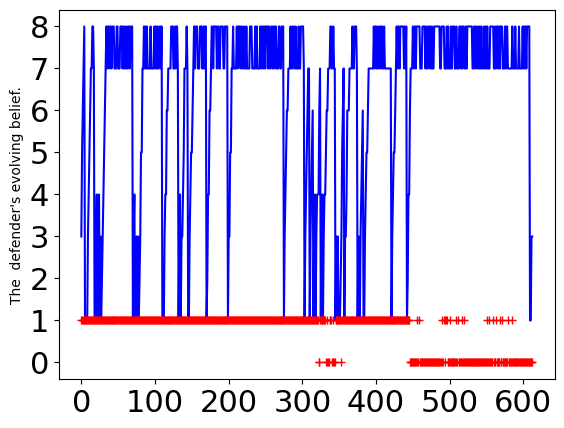}
		\caption{Sensor (b),B,I}
	\end{subfigure}
	\begin{subfigure}[b]{0.45\textwidth}
	\includegraphics[width= \textwidth]{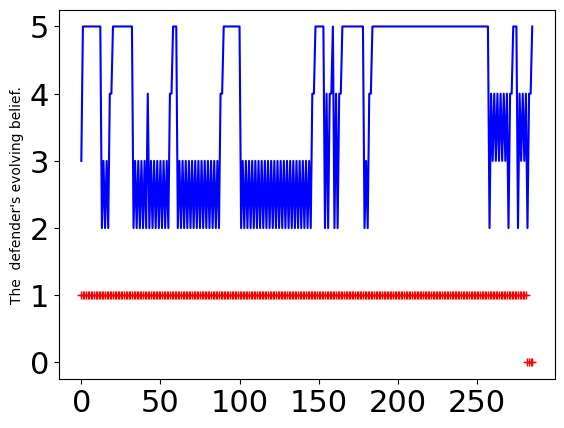}
	\caption{Sensor (c),B,I}
\end{subfigure}
\begin{subfigure}[b]{0.45\textwidth}
	\includegraphics[width= \textwidth]{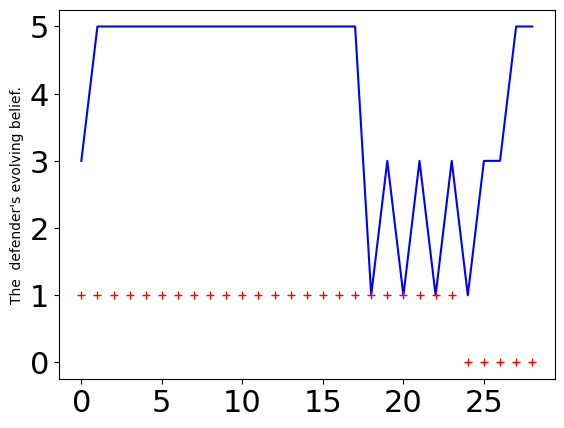}
	\caption{Sensor (c),P,I}
\end{subfigure}
	%     \begin{subfigure}[b]{0.3\textwidth}
		%    \includegraphics[width=0.3\textwidth]{belief_update_ex4}
		%\end{subfigure}
		\caption{The evolving belief of the defender over time  in a single run under the attacker's \ac{asw} deception strategy for different sensor configurations. }
		\label{fig:belief_update}
		\end{figure}

 Note that the attack strategy outputs a set of actions to be taken at each augmented state. An \ac{asw} strategy only needs to take each action with a nonzero probability to ensure the objective is satisfied with probability one. It is possible to compute a non-revealing, intention deception \ac{asw} strategy that minimizes the number of time steps for the attacker to reach the goal.  This is solved using a stochastic shortest path algorithms in the augmented \ac{mdp} where  the attacker's actions are restricted to actions allowed by the deception \ac{asw} strategy.
 Next, we %use the stochastic shortest path algorithm to compute an optimal, non-revealing, intention deception \ac{asw} strategy that minimizes the number of time steps for the attacker to reach the goal. T
 exercise the optimal attack strategy in the the game with different sensor configurations to obtain sampled winning runs for the attacker. 
 In Fig.~\ref{fig:belief_update}, the blue lines describes the size of the defender's belief over time and the red crosses represents if the current state is in the belief of the defender (1 for yes and 0 for no). Comparing $\mbox{(a)-B-I}$ to $\mbox{(b)-B-I}$,  it is observed that the defender can reduce the uncertainty in his belief more frequently in the case of $\mbox{(a)-B-I}$ and the attacker has to spend a longer time in order to satisfy the attack objective with deceptive strategy. In the case of $\mbox{(b)-B-I}$, the defender is uncertain of the attacker's state for a majority of time steps.  When it comes to the sensor configuration $\mbox{(c)-B-I},\mbox{(c)-P-I}$, we observe that the attacker achieves the objective in a much shorter time comparing to the previous cases.  
As expected, prior to reaching the attack goal states $4$, the defender's belief  no longer contains  the true state \footnote{Videos of the sampled runs for case $\mbox{(c)-B-I},\mbox{(c)-P-I}$ can be found  at \url{https://bit.ly/3BiPRb9} where the light green cells are states in the defender's belief.}.

Given this analysis, the synthesized intention deception attack strategy can be used to effectively evaluate the blind spots of a security monitoring system. It is also possible to use that attack strategy as a counterexample for counter-example guided security system design.

%The common feature in all these cases is that the attacker can reach the attack goal state only after that the defender's belief no longer contains the true state. 

\begin{table}[ht!]
	\caption{Experiment results with different sensor configurations.  \label{tbl:results}}
	\begin{tabular}{c|c|c|c|c}
		Sensor config.  & Win initial states & Computation time & Size of augmented \ac{mdp} & $\abs{\asw}$  \\
		\hline
		(a)-B-I& $[20 .. 24]$ &0.233& 1091 & 374\\
		(b)-B-I& $5,10,15$, $[20..24]$ & 0.4086 & 1339 & 682\\
		(c)-B-I & $5,10,15$, $[20..24]$ & 0.0569 & 497&340\\
		(c)-P-I & $[20.. 24]$ &  0.0415 & 359& 170 \\
		(c)-P-V & $\emptyset$ & 0.0562 & 519 &20\\
		\hline
	\end{tabular}
 \end{table}

%% file: sensor_scheduler.tex
\begin{tikzpicture}[->,>=stealth',shorten >=1pt,auto,node distance=2cm,
        scale = 0.6,transform shape]

  \node[state,initial] (1) {$1$};
  \node[state] (2) [right of=1] {$2$};
  \node[state] (3) [below of=2] {$3$};
  \node[state] (4) [below of=1] {$4$};

  \path (1) edge              node {$0.5$} (2)
        (1) edge [loop above]             node {$0.5$} (1)
        (2) edge [loop right]             node {$0.5$} (2)
        (2) edge              node {$0.5$} (3)
        (3) edge  [loop right]            node {$0.5$} (3)
        (3) edge              node {$0.5$} (4)
        (4) edge              node {$0.5$} (1)
        (4) edge   [loop left]           node {$0.5$} (4);

\end{tikzpicture}

%% file: appendix.tex
\appendix

\section{Proof of Proposition~\ref{prop:property-of-ASW} and the construction of \ac{asw} region.}
\begin{proof}
	First, we provide the algorithm to solve the \ac{asw} region $\asw(\varphi)$ where $\varphi \coloneqq \neg U \until F$ where $U \cap F = \emptyset$.
	\begin{enumerate}
		\item Initiate $X_0=F$ and $Y_0=S \setminus U$. Let $i=j=1$.
		\item Let $X_{i+1} =  \{X_i\} \cup \{s \in Y_j \setminus X_i\mid \exists a\in A(s), \post(s,a) \cap X_i \ne \emptyset \text{ and } \post(s,a)\subseteq Y_j\}$ 
		\item If  $X_{i+1}\ne X_i$, then let $i=i+1$ and go to step 2; else, let $n=i$ and go to step 4.
		\item Let $Y_{j+1}=X_i$. If $Y_{j+1}= Y_j$, then $\asw(\varphi)=Y_j$. Return $\{X_i, i=1,\ldots, n\}$ computed from the last iteration. Else, let $j=j+1$ and $i=0$ and go to step 2. 
	\end{enumerate}
	The algorithm returns a set of level sets $X_i, i=0,\ldots, n$ for some $n\ge 0$ and the \ac{asw} region $\asw(\varphi)$. Recall that 
	$\allowed: S\rightarrow 2^A $ is defined by $\allowed(s) = \{a\in A(s)\mid \post(s,a)\subseteq \asw(\varphi)\}$.
	The following property holds: For each $s\in X_i\setminus X_{i-1}$, there exists an action $a\in \allowed(s)$ that ensures, with a positive probability, the next state is in $X_{i-1}$ and with   probability one, the next state is in $\asw(\varphi)$.

	Let's define a function $\prog: \asw(\varphi)\rightarrow 2^A$ such that for each $s\in X_i\setminus X_{i-1}$, $\prog(s)=\{a\in \allowed(s)\mid   \post(s,a)\cap X_{i-1}\ne \emptyset\}$. Intuitively, the set $\prog(s)$ is a set of actions, each of which ensures that a progress (to a lower level set) can be made with a positive probability. 
	
	Therefore, by following a policy $\pi$ that selects any action in $\allowed(s)$ with probability $> 0$, 
	the probability of starting from a state $s\in X_i\setminus X_{i-1}$ and  not reaching a state in $X_0=F$ in $i$ steps is less than $(1-p)^i$ where $p = \min_{0<i \le n, s\in S, a\in\prog(s)} \pi(s,a)P(s' \mid s,a)$ and is nonzero. If in the $i$-th step, the set $X_0$ is not reached, the agent will reach a state $s'\in \asw(\varphi)$ from which an action  	in $\prog(s')$ will be selected with a nonzero probability. Thus, the probability of never reaching a state in $F$ is $\lim_{k\rightarrow \infty} (1-p)^k =0$. In other words, the policy $\pi$ ensure $F$ is eventually reached with probability one. At the same time, because the set $Y_j \cap U =\emptyset$ for all $j >0$ during iterations, $\asw(\varphi)\cap U = \emptyset$ and thus the probability of reaching a state in $U$ is zero by following the policy $\pi$.
\end{proof}

\section{Proof of Theorem~\ref{thm:action-invisible}}

\begin{proof}
	
	We show that the \ac{asw} policy $\widehat{\pi}_1: S\times 2^S\rightarrow \dist{A}$ obtained from the augmented \ac{mdp} is qualitatively observation-equivalent to an \ac{asw} policy for the user.
	
	Consider a history $h  = s_0a_0s_1a_1\ldots s_n $ which is sampled from the stochastic process $M_{\widehat \pi_1}$ and satisfies $s_i \notin F_1 \cup U_1$ for $0\le i <n$ and  $s_n \in F_1$ . The history is associated with a history in the augmented \ac{mdp}, $\widehat h =  (s_0, B_0)a_0(s_1,B_1)a_1\ldots (s_n,B_n)$  where $B_0=\obs_S(s_0)$ is the initial belief for the defender. Due to the construction of the augmented \ac{mdp}, for all $0\le i \le n$, $B_i\ne \emptyset$.

	By the definition of qualitatively observation-equivalence, we only need to show that there exists $h' =s_0'a'_0s_1'a'_1 \ldots s_n'$ where $s_i'\in B_i$, for all $i=0,\ldots, n$, such that $\Pr(h', M_{\pi_0})>0$ where $M_{\pi_0}$ is the Markov chain induced by a user's \ac{asw} policy $\pi_0$ from the \ac{mdp} $M$. It is observed that $a_i$ and $a'_i$ may not be the same. 
		By way of contradiction, suppose, for any state-action sequence 
	$h'  =s_0'a'_0s_1'a'_1 \ldots s_n' $ where   $s_i'\in B_i$ and $a'_i \in A(s_i')$ for $0\le i \le n$, it holds that $\Pr(h', M_{\pi_0})=0$. When $\Pr(h', M_{\pi_0})=0$, there are two possible cases: First case:  there exists some $i\ge 0$ such that for all $s\in B_i$, $ \allowed(s)=\emptyset$; second case: there does not exists an action $a$ enabled from the belief $B_i$ and a state $s'\in B_{i+1}$ such that $P(s'|s,a)>0$.
	%\aknote{See if $\obs_S$ partitioning $S$ has any effect on this argument.}
	
	The first case is not possible because when for all $s\in B_i$,  $\allowed(s)=\emptyset$,  then the next state reached will be $(s_i,\emptyset)$ which is  a sink state, contradicting the fact that $\widehat h
	$ satisfies the reach-avoid objective. In the second case,  if for any state $s\in B_i$, for any action $a$  enabled from $s$, $\post(s,a) \cap B_{i+1}=\emptyset$, then $B_{i+1}=\emptyset$, this again contracts that $\widehat h$ visits a state in $\widehat F$.
	
	Thus, it holds that there exists $h'$  such that $\obs_S(h)=\obs_S(h')$ and $\Pr(h', M_{\pi_0})>0$.  
\end{proof}